\parindent=0pt
\parskip=0.6cm
\magnification \magstephalf

\null
\vskip 1.5cm
\centerline{{\bf DECELERATING UNIVERSES OLDER THAN THEIR HUBBLE TIMES}
\footnote{\dag}{Mon. Not. R. Astron. Soc. 297 (1998) 923}}
\vskip 1.5cm

\centerline {by}
\vskip 0.3cm

\centerline {J.C. Jackson\footnote{*}{e-mail: john.jackson@unn.ac.uk} and Marina Dodgson}
\centerline {Department of Mathematics and Statistics}
\centerline {University of Northumbria at Newcastle}
\centerline {Ellison Building}
\centerline {Newcastle upon Tyne NE1 8ST, UK}
\vskip 1.5cm

\centerline{\bf ABSTRACT}
\vskip 0.3cm

Recent observations suggest that Hubble's constant is large, and hence that 
the Universe appears to be younger than some of its constituents. 
The traditional escape route, which assumes that the expansion 
is accelerating, appears to be blocked by observations of Type 1a supernovae, 
which suggest that the Universe is decelerating. These observations are 
reconciled in a model in which the Universe has experienced an inflationary 
phase in the recent past, driven by an ultra-light inflaton whose Compton 
wavelength is of the same order as the Hubble radius.
\vskip 0.6cm

{\bf Key words:} cosmology -- observations -- theory -- dark matter. 
\vfil\eject

{\bf 1 INTRODUCTION}

From time to time, cosmologists have had to face the possibility 
that the Hubble time $1/H_0$ is shorter than the ages of the oldest 
Galactic globular clusters (the so-called age crisis), and hence that 
the Universe appears to contain objects older than itself.  This 
certainly seemed to be the case when determinations 
of Hubble's constant using extra-galactic Cepheids (Pierce et al. 1994; 
Freedman et al. 1994; Tanvir et al. 1995) produced values of 
$80\pm 17$, $87\pm 7$ and $69\pm 8$ km sec$^{-1}$ Mpc$^{-1}$ respectively, 
with a best estimate based upon these figures of $79\pm 5$
km sec$^{-1}$ Mpc$^{-1}$ (see also van den Bergh 1995).  However, 
subsequent recalibration of the Cepheid period-luminosity zero-point, 
using data from the Hipparcos astrometry satellite (Feast \& Catchpole 1997), 
has indicated that these figures should be reduced by 10\%, giving 
$H_0=71.3\pm 4.5$ km sec$^{-1}$ Mpc$^{-1}$; the corresponding Hubble 
time is $t_{H_0}=13.8\pm 0.9$ Gyr (see, however, Madore \& Freedman 1997).

The existence of globular clusters older than 15 Gyr was believed 
to be well established (see for example Sandage 1982), but more recent 
re-evaluations have given ages of $15.8\pm 2.0$ Gyr based on main-sequence 
turnoff (Bolte \& Hogan 1995), and $13.5\pm 2.0$ Gyr based on giant-branch 
fitting (Jimenez et al. 1996), with a best estimate based upon these figures 
of $14.7\pm 1.4$ Gyr (see also Chaboyer at al. 1996).  
However, the aforementioned Cepheid recalibration has implications here too,
via a corresponding recalibration of RR Lyrae distances to these clusters, 
which indicates for example that the main-sequence turnoff stars are 
brighter and younger than previously supposed; Feast \& Catchpole (1997) 
suggest an age reduction of 23\%.  In similar vein, cluster distances
established by main-sequence fitting, to a sample of local subdwarfs with
Hipparcos parallaxes, has yielded ages between $12$ and $13$ Gyr 
(Gratton et al. 1997; Reid 1997).  

On the basis of these revised figures it is tempting to say that there is 
no age crisis, but we are not convinced that this is so.  A formation time 
must be added to cluster ages, and there is abundant evidence that the 
Universe contains some dark matter, enough to reduce its age significantly 
compared with the Hubble time, for example $t_0\leq 0.845t_{H_0}$ if 
$\Omega_0\geq 0.2$, in which case our best estimate of $H_0$ gives 
$t_0\leq 11.7\pm 0.8$ Gyr.  In any case, low cluster ages are not indicated 
by all of the most recent data.  Renzini et al. (1996) have presented 
Hubble Space Telescope observations of white dwarfs in the local 
neighbourhood and in the cluster NGC 6752; matching the two cooling 
sequences yields a distance modulus for the latter, and an age of 
$15.0\pm 1.5$ Gyr (however, Hipparcos paralaxes have yet to be used in 
this context).  Finally thorium nucleocosmochronometry applied to a 
metal-poor halo star (Cowan et al. 1997) has yielded an age of $17\pm 4$ Gyr.

If there is an age crisis, the way out is to suppose that the Universe 
has passed through an accelerating phase, traditionally effected 
by a positive cosmological constant $\lambda$, which allows $t_0>t_{H_0}$;  
particularly fashionable have been those models which maintain spatial 
flatness by assuming the appropriate admixture of material and vacuum energy,
$\Omega_0+\Lambda_0=1$ (where $\Lambda_0=\lambda/3H_0^2$), an idea due 
originally to Peebles (1984); as a precautionary measure this escape 
route has been reopened (Ostriker \& Steinhardt 1995).  The problem with 
such a scenario is that it implies that the expansion of the Universe now 
should be seen to be accelerating, whereas very recent evidence, in the form 
of a traditional Hubble diagram based upon Type 1a supernovae, suggests that 
this is not the case, and that the deceleration parameter satisfies 
$0.14<q_0<0.79$, with a central value of $0.47$ (Perlmutter et al. 
1996, 1997a,b); the precise figures (but not the sign) depend upon the 
value of $\lambda$, here taken to be zero.  Of course the classical Hubble 
diagram based upon first-ranked cluster galaxies always did suggest a 
positive value for $q_0$ (Sandage 1988; Sandage, Kron \& Longair 1995), 
but faith in this conclusion was never very strong, owing to the uncertain 
role of evolutionary effects; the case for a non-negative deceleration 
parameter now seems much stronger, and is one of the prime factors 
motivating this work.

It is certainly not our intention to insist that $q_0>0$, and that the 
oldest stars have ages which are longer than the Hubble time, but at
face value the observations allow both statements as more than a 
possibility, and to insist that one of them must be wrong (McGaugh 1996) 
is to risk falling into a teleological trap. We feel that the 
addition of an appropriate model to the cosmological compendium is long  
overdue, that is one which has decelerating phases in which $t_0$ is greater 
than $t_{H_0}$; the purpose of this work is to examine what reasonable 
options might be on offer in this context.  In Section 2 we discuss models 
which are dominated by a rolling homogeneous scalar field, and in the
Section 3 we confirm that such a model can account for the supernova 
observations reported by Perlmutter et al.
\vskip 0.6cm

{\bf 2 LATE-STAGE INFLATION}

A suitable model must retain some of the features of those discussed by
Peebles (1984) and Ostriker \& Steinhardt (1995), but with an added degree 
of flexibility.  Models in which the current epoch is dominated by a 
homogeneous scalar field have been considered before in the context of 
timescales and missing matter, by Olson \& Jordan (1987), Peebles \& Ratra 
(1988), and particularly Ratra \& Peebles (1988), who introduce explicitly 
the idea of an ultra-light boson as a CDM candidate; this possibility has 
been rediscovered recently by Frieman et al. (1995), who examine 
how such a boson might arise naturally, according to modern theories of 
fields and particles.  The model which we have adopted is based upon a 
simplified version of their ideas.  Although much of what we have to say is 
implicit in earlier work, to our knowledge the key points which we wish to 
bring out have not been made before.  The basic idea is to adopt the 
inflationary model of the very early Universe, adapted to describe recent 
cosmological history by rescaling the appropriate length/mass scales by a 
factor of approximately $10^{60}$.

The idea that the dynamics of the very early Universe were dominated by a 
homogeneous isotropic scalar field, with effective classical Lagrangian 
${\cal L}=\dot\phi^2/2-V(\phi,T)$, has dominated cosmological
thinking for almost two decades.  In so-called new inflation (Linde 1982), 
there is an initial period in which the Lagrangian is dominated by the 
potential term $V(\phi,T)$, during which the Universe inflates and cools 
below a transition temperature $T_c$; after this event the field $\phi$ 
begins to move from its false-vacuum value of zero, towards a non-zero value 
$\phi_0$ at which $V$ has a minimum, about which value $\phi$ undergoes 
decaying oscillations, effected by dissipation due to coupling to other 
matter fields.  As originally conceived the inflationary process finishes 
after a time of typically several hundred Planck times, and is thus shielded 
for ever from the possible embarassment of direct observation.  
However, this need not be the case; a trivial example would be when the true 
vacuum (minimum) value of $V$ is not exactly zero, which would constitute a 
cosmolgical constant $\lambda$ whose effects might be observable now.  
More generally, we might expand $V$ as a Taylor series about $\phi_0$:
$$
V(\phi)=V(\phi_0)+{1\over 2}V''(\phi_0)(\phi-\phi_0)^2+...
\eqno(1)
$$
For $\phi$ close to $\phi_0$, we retain just these terms and move the
origin of $\phi$ to $\phi_0$, giving
$$
V(\phi)={\lambda\over 8\pi G}+{1\over 2}\omega_c^2\phi^2
\eqno(2)
$$
where $\omega_c$ is the Compton freqency of the associated Nambu-Goldstone
boson.  In chaotic inflation (Linde 1983) the potential has the form (2) 
{\it ab initio}, and inflation is attributed to initial conditions 
$\omega_c \phi\gg \dot\phi$ over a suitably large patch of the primordial 
chaos.  The mass $\hbar\omega_c$ is usually very large, typically of order 
$10^9$ Gev, with corresponding freqency $\omega_c\sim 10^{34}$ sec$^{-1}$.  
Here we examine the possibility that the current dynamics of the Universe 
are dominated by a scalar field corresponding to a very light boson, 
with mass such that $\omega_c$ and $H_0$ are of the same order.

The basic scene is thus a re-run of chaotic inflation, 
in which the scalar field mimics a positive cosmological constant 
in a slow-roll phase until the Hubble time exceeds $\omega_c^{-1}$, 
after which it undergoes damped oscillations (that is Hubble damping; 
we assume that the scalar field does not couple to ordinary matter).  
It is during the period immediately after inflation that such a 
model would exhibit the behaviour we are seeking to illustrate.  
Corresponding to the potential (2), we have effective density 
$\rho_\phi=(\dot\phi^2+\omega_c^2\phi^2)/2$ and pressure 
$p_\phi=(\dot\phi^2-\omega_c^2\phi^2)/2$, and the Friedmann equations 
for the scale factor $R(t)$ become
$$
\ddot R=-{4\pi \over 3}G(\rho_\phi+3p_\phi+\rho_m)R=-{4\pi \over 3}G(2\dot\phi^2-\omega_c^2\phi^2+\rho_m)R,
\eqno(3)
$$
$$
\dot R ^2={8\pi \over 3}G(\rho_\phi+\rho_m)R^2-kc^2,
\eqno(4)
$$
where we have introduced ordinary pressure-free matter with density $\rho_m$
and set the true cosmological constant $\lambda=0$.  The scalar field is
governed by the equation
$$
\ddot\phi+3H\dot\phi+\omega_c^2\phi=0
\eqno(5)
$$
where $H=\dot R/R$ is Hubble's `constant' (see for example Peebles 1993).
We introduce the usual density parameter $\Omega_m=8\pi G\rho_m/3H^2$,
and similar parameters associated with the scalar field, 
$\Omega_{\rho_\phi}=8\pi G\rho_\phi/3H^2$ and
$\Omega_{p_\phi}=8\pi Gp_\phi/3H^2$.  Equation (3) and (4) then give the 
deceleration and curvature parameters as 
$$
q=(\Omega_{\rho_\phi}+3\Omega_{p_\phi}+\Omega_m)/2
\eqno(6)
$$
and
$$
K=\Omega_{\rho_\phi}+\Omega_m-1.
\eqno(7)
$$
To solve the system of equations (3) to (5), we introduce variables
$u=\sqrt{G}\phi$, $v=\dot u$, $x=R/R(0)$, $y=\dot x$, and measure time 
in units of $\omega_c^{-1}$, when these equations are re-cast in the form
of an autonomous non-linear dynamical system:
$$\eqalignno{
\dot u&=v,~~~~~~\dot v=-3yv/x-u,&(8)\cr
\dot x&=y,~~~~~~\dot y=-4\pi x(2v^2-u^2)/3-\textstyle{1 \over 2}\omega_m(0)/x^2&(9)\cr
}$$
where $\omega_m=8\pi G\rho_m/3\omega_c^2$.  At some arbitrary time $t=0$ 
we specify initial values $x(0)=1$, $H(0)$, $\Omega_m(0)$, 
$\Omega_{\rho_\phi}(0)$, and $\Omega_{p_\phi}(0)=-\Omega_{\rho_\phi}(0)$, 
the latter ensuring that the initial value of $\dot\phi=0$,
and that the initial effects of the scalar field are inflationary.
At the moment an exhaustive survey of this parameter space would be 
premature; we shall be content to find a representative model which is 
compatible with the several observational criteria established in Section 1.

We choose to concentrate on flat models, in part for mathematical 
convenience, and in part because the dictates of orthodox inflation 
still carry some weight.  This does not represent a prejudice against 
open models ($k=-1$) on our part; the case for an open Universe is 
in fact quite strong (see for example Coles \& Ellis 1994).  A convenient 
starting point is $\Omega_m(0)=0.999$, $\Omega_{\rho_\phi}(0)=0.001$ 
(i.e. $K=0$ according to equation (7)), and we have a one-parameter set 
of models depending upon $H(0)$ (equivalent to a choice of $\omega_c$).  
We first run the system backwards to locate the initial singularity, 
and then forwards to locate the transition from the early matter-dominated 
phase to the final state in which the scalar field is dominant.  
Figure 1 shows the scale-factor $x(t)$ for a typical run, with $H(0)=27.2$, 
showing point A marking the first switch from acceleration to deceleration, 
and the region A to C where the tangent hits the positive $t$-axis, 
in other words points in this region are older than the corresponding Hubble 
time.  This is quantified in Figure 2, in which $t-t_H$ (continuous) and 
$q$ (dashed) are plotted against $t$; thus if `now' (point B of Figure 1)
corresponds to $q=0.5$, we see that $t_0=2.81$ exceeds the current Hubble 
time by 8\%.  Figure 3 shows $\Omega_{\rho_\phi}$ (continuous), 
$\Omega_m$ (dashed), and $\Omega_{\rho_\phi}+3\Omega_{p_\phi}$ (dash-dotted), 
the effective gravitational density of the scalar field; it is the latter's 
oscillation between positive and negative values which accounts for the way 
in which $q$ changes sign.  The matter/scalar field transition at $t=0.73$ 
is quite clear, corresponding to a redshift $z=3.1$; $\Omega_m$ eventually 
settles down to a value of 0.10.  An element of fine tuning is required to 
achieve acceptable values of the asymptotic value of $\Omega_m$, but not too 
fine; $H(0)>40$ results in less matter than we see ($\Omega_m<0.003$), 
whereas $H(0)<23$ results in too much dark matter ($\Omega_m>0.2$), in the 
sense that an epoch in which $t>t_H$ is precluded.  There are also open and 
closed models which exhibit similar behaviour, including some with long 
Lema\^\i tre-like coasting phases.  The fixed non-zero asymptotic values 
of $\Omega_m$ are a particularly attractive feature of our flat models, 
compared with those with $\Omega_0+\Lambda_0=1$, in which $\Omega_m\sim 0.1$ 
is but a passing phase. 

\vskip 0.6cm

{\bf 3 COMPATIBILITY WITH OBSERVATIONS}

A putative cosmological model must expect to run the gauntlet of an
increasing number of neo-classical observational tests ($z{}^{<}_\sim 4$), 
and of modern tests based upon structure formation and the cosmic microwave
background ($z{}^{<}_\sim 1000$), but in the context of this model 
a comprehensive survey would seem to be premature until such time as the age 
crisis has a firmer observational basis.  As our universe combines aspects 
of several models which separately have their advocates, gross contradictions
are unlikely.  Nevertheless, our ideas will not have achieved their objective 
if the corresponding magnitude--redshift curves are incompatible 
with the observations described by Perlmutter et al.; we concentrate on the 
flat model examined in Section 2, and on an observer at point B of Figure 1, 
where $q_0=0.5$.  It is well-known that the initial (low $z$) shape 
of the $m-z$ relationship is determined by the value of $q_0$, and not 
by the details of how this deceleration is produced (see for example 
Weinberg 1972), so that compatibility with the preferred matter-dominated 
model ($q_0=0.47$) is almost guaranteed.  However, as the largest redshift 
(0.458) in the sample is somewhat larger than allowed by the low--$z$ 
approximation, we must look at the matter in more detail, particularly 
to predict the divergences which should be observed at higher redshifts.  

Computation of the appropriate $m-z$ curve is quite straightforward within 
the framework developed in Section 2; we read off the parameters 
$\Omega_m(B)$, $\Omega_{\rho_\phi}(B)$, $x(B)$ and $H(B)$, which are deemed 
to be current values; $x$ and $t$ are then rescaled, $x\rightarrow x/x(B)$, 
$t\rightarrow H(B)t$, and for the spatially flat model in question, 
standard lore then gives the luminosity distance as
$$
d_L(t(z))={c \over H_0}\int_{t(z)}^{t_0}{dt \over x(t)}.
\eqno(10)
$$ 
The integration is performed numerically, over the table of $x(t)$ 
values established in Section 2, with $z(t)=1/x(t)-1$.  A specific
value for $\omega_c$ can be derived by comparing $H(B)=0.380$, which 
is Hubble's constant in terms of the Compton frequency, with $H_0$ in 
conventional units, giving $\omega_c=H_0/H(B)=8.53h\times 10^{-18}$ 
sec$^{-1}$, and a corresponding mass of $5.61h\times 10^{-33}$ ev 
($h=1\Rightarrow H_0=100$ km sec$^{-1}$ Mpc$^{-1}$).  The corresponding 
$m-z$ curve is the continuous one in Figure 4, to be compared with the dashed 
curve, which is the canonical $\Omega_0=1$, $q_0=0.5$ matter-dominated 
case.  The curves are virtually indistinguishable until $z$ exceeds $0.5$,
after which the divergence becomes quite pronounced; if these supernovae 
fulfill their promise as standard candles, this divergence would be 
detectable when objects in the redshift range 0.5 to 1.0 are observed. 

Figure 4 also shows (dotted) a popular flat accelerating model, 
$\Omega_0=0.1$ (i.e. the asymptotic value in our scalar universe) and 
$\Lambda_0=0.9$; as would be expected the scalar curve veers towards 
the accelerating case at high redshifts.  We have examined this effect in 
the context of a neo-classical test which involves much higher redshifts 
than those associated with supernovae, namely gravitational lensing of
quasars by intervening galaxies.  The integrated probability $\tau(z)$ 
that a quasar at redshift $z$ is multiply imaged by a galaxy along the 
line of sight (the gravitational lensing optical depth) is particularly 
simple in spatially flat universes, being given by
$$
\tau(z)={F \over 30}d_M(z)^3,
$$
where $d_M(z)=d_L(z)/(1+z)$ is the so-called proper-motion distance 
(Weinberg 1972); $F$ is a dimensionless measure of the lensing
effectiveness of the galaxies, here represented by a non-evolving 
population of randomly-distributed singular isothermal spheres
(for further details see Turner, Ostriker \& Gott 1984; 
Gott, Park \& Lee 1989; Turner 1990).  Thus our work on the Hubble 
diagram enables curves of $\tau(z)/F$ to be produced easily, and the 
three cases considered above are shown in Figure 5.  As is well-known, 
the lensing optical depth in the $\lambda$-dominated case is typically
an order of magnitude larger than that in the matter-dominated one; 
the scalar model falls between the two.

The observational situation is very confusing; until 1993
the frequency of lensing events was believed to be very low,
with a sample of 402 quasars showing no such events (Boyle,
Fong, Shanks \& Peterson 1990), and one comprising 4250 objects
showing at most 9 examples (Hewitt \& Burbidge 1987, 1989).
On the basis of these figures, and an estimate of the lensing strength 
$F=0.15$, Turner (1990) concluded that among the conventional $k=0$, 
$\Omega_0+\Lambda_0=1$ models, those with $\Omega_0\sim 1$ were very 
much favoured.  A weakened version of this conclusion survived a downward 
revision of the lensing strength, to $F=0.047$ (Fukugita \& Turner 1991;
Fukugita et al. 1992), which authors also considered selection biases due to 
finite angular resolution and magnification, which can be large but 
fortunately have opposing effects and tend to cancel.  However, the advent 
of the Hubble Space Telescope Snapshot survey (Bahcall et al. 1992;
Maoz et al. 1992, 1993a,b, hereafter referred to generically as Maoz et al.) 
has radically changed the situation; this survey reports a lensing frequency 
almost a factor of 10 higher than earlier ones, 3 to 6 examples in a sample 
of 502 quasars, and in this context at least low $\Omega_0$--high $\Lambda_0$ 
models appear to be supported (see also Chiba \& Yoshii 1997).  The 
discrepancy is believed to be due to inadequate angular resolution in the 
case of pre-HST surveys.  We shall compare predicted lensing frequencies
with the HST survey, using Fukugita \& Turner's value $F=0.047$. 
We assume following Maoz et al. that the HST angular resolution is 
sufficient to preclude any selection bias against small angular separations; 
however, the tendency of lensed quasars to be over-represented in a 
magnitude-limited sample due to their increased apparent brightness must 
be allowed for, and we adopt a fixed magnification-bias factor of 2.5, as 
suggested by Fukugita \& Turner for deep surveys.  Opinion is now fairly 
firm (Maoz \& Rix 1993; Torres \& Waga 1996) that of the six lensing events 
mentioned by Maoz et al., one is spurious and another is due to an
intervening cluster of galaxies, rather than a single galaxy; thus we take
4 out of 502 as the observed lensing frequency.  The predicted frequencies 
for the canonical CDM, scalar and $\lambda$ models are respectively 
1.3, 3.9 and 8.8, with respective binomial probabilities of 0.03, 0.196 
and 0.04.  It is gratifying to see our model performing so well in this 
context, but in view of the various uncertainties we cannot regard these 
results as anything more than illustrative.
\vskip 0.6cm

{\bf 4 CONCLUSIONS}

Radical measures are contemplated here, of seemingly epicyclic proportions,
but there is a growing body of observational evidence that such measures
might be called for.  (Indeed as this conclusion is being written we note
that the $H_0$ pendulum is swinging back towards higher values,
according to a re-evaluation of the extragalactic Cepheid distance
scale in which metallicity is considered (Kochanek 1997); figures 
from $69$ km sec$^{-1}$ Mpc$^{-1}$ to $90$ km sec$^{-1}$ Mpc$^{-1}$
are mentioned.)  These measures are based upon a simple and natural
generalization of the concept of vacuum energy, represented by equation (2), 
which we believe to be the minimal form necessary, to account for the 
putative observational paradox which has motivated this work. We 
have considered two neo-classical cosmological tests, with regard to which 
the behaviour of our scalar universe is satisfactory, and falls between the 
two extremes defined by the standard matter-dominated model and a 
$\lambda$-dominated one; in this context we believe this to be generally true.
\vfil\eject

{\bf REFERENCES}
\vskip 0.3cm

{\obeylines\parskip=0pt

Bahcall J.N., Maoz D., Doxsey R., Schneider D.P., Bahcall N.A., Lahav O., Yanny B., 
\qquad 1992, ApJ, 387, 56
Bolte M., Hogan C.J., 1995, Nat, 376, 399
Boyle B.J., Fong R., Shanks T., Peterson B.A., 1990, MNRAS, 243, 1

Chaboyer B., Demarque P., Kernan P.J., Krauss L.M., 1996, Sci, 271, 957
Chiba M., Yoshii Y., 1997, 489, 485
Coles P., Ellis G.F.R., 1994, Nat, 370, 609
Cowan J.J., McWilliam A., Sneden C., Burris D.L., 1997, ApJ, 480, 246

Feast M.W., Catchpole R.M., 1997, MNRAS, 286, L1
Freedman, W.L. et al., 1994, Nat, 371, 757 
Frieman J.A., Hill C.T., Stebbins A., Waga I., 1995, Phys. Rev. Lett., 75, 2077
Fukugita M., Turner E.L., 1991, MNRAS, 253, 99
Fukugita M., Futamase T., Kasai M., Turner E.L., 1992, ApJ, 393, 3

Gratton R.G., Fusi Pecci F., Carretta E., Clementi G., Corsi C.E., Lattanzi M.G., 1997,
\qquad ApJ, 491, 749
Gott J.R., Park M.G., Lee H.M., 1989, ApJ, 338, 1

Hamuy  M., Phillips M.M., Maza J., Suntzeff N.B., Schommer R., Aviles R., 1995, 
\qquad AJ, 109, 1

Hewitt A., Burbidge G., 1987, ApJS, 63, 1
Hewitt A., Burbidge G., 1989, ApJS, 69, 1

Jimenez R., Thejjl P., Jorgensen U.G., MacDonald J., Pagel B., 1996, MNRAS, 282, 926

Kochanek C.S., 1997, ApJ, 491, 13

Linde A.D., 1982, Phys. Lett., B108, 389
Linde A.D., 1983, Phys. Lett., B129, 177

Madore B.F., Freedman W.L., 1998, ApJ, 492, 110
Maoz D., Rix H.-W., 1993, ApJ, 416, 425
Maoz D., Bahcall J.N., Schneider D.P., Doxsey R., Bahcall N.A., Lahav O., Yanny B., 1992,     
\qquad ApJ, 394, 51
Maoz D., Bahcall J.N., Schneider D.P., Doxsey R., Bahcall N.A., Lahav O., Yanny B., 1993a, 
\qquad ApJ, 402, 69
Maoz D. et al., 1993b, ApJ, 409, 28
McGaugh S.S., 1996, Nat, 381, 483 

Olson T.S., Jordan T.F., 1987, Phys. Rev., D35, 3258
Ostriker J.P., Steinhardt P.J., 1995, Nat, 377, 600

Peebles P.J.E., 1984, ApJ, 284, 439
Peebles P.J.E., 1993, Principles of Physical Cosmology.  Princeton University Press, pp. 394-396
Peebles P.J.E., Ratra, B., 1988, ApJ, 325, L17
Perlmutter S. et al., 1996, Nucl. Phys., S51B, 20
Perlmutter S. et al., 1997a, in Canal R., Ruiz-LaPuente P., Isern J., eds., 
\qquad Thermonuclear Supernovae.  Kluwer, Dordrecht, pp. 749-763
Perlmutter S. et al., 1997b, ApJ, 483, 565
Pierce M.J., Welch D.L., McClure R.D., van den Bergh S., Racine R., Stetson P.B.,
\qquad 1994, Nat, 371, 385

Ratra B., Peebles P.J.E., 1988, Phys. Rev., D37, 3406
Reid I.N., 1997, AJ, 114, 161
Renzini A. et al., 1996, ApJ, 465, L23

Sandage A., 1982, ApJ, 252, 553
Sandage A., 1988, ARA\&A, 26, 561
Sandage A., Kron R.G., Longair M.S., 1995, The Deep Universe, 
\qquad SAAS-FEE Advanced Course 23, Springer, Berlin

Tanvir N.R., Shanks, T., Ferguson H.C., Robinson D.R.T., 1995, Nat, 377, 27
Torres L.F.B., Waga I., 1996, MNRAS, 279, 712
Turner E.L., 1990, ApJ, 365, L43
Turner E.L., Ostriker J.P., Gott J.R., 1984, ApJ, 284, 1

van den Bergh S., 1995, Sci, 270, 1942

Weinberg S., 1972, Gravitation and Cosmology.  Wiley, New York, pp. 423 \& 442
}
\vskip 0.6cm

{\bf ACKNOWLEDGMENTS}

Marina Dodgson acknowledges receipt of an UNN internal research studentship,  
1993-96, during the tenure of which this work was initiated.  It is a 
pleasure to thank Richard McMahon, of the Institute of Astronomy, University
of Cambridge, for advice about the rapidly developing observational scene.
\vskip 0.6cm

{\bf FIGURE CAPTIONS}

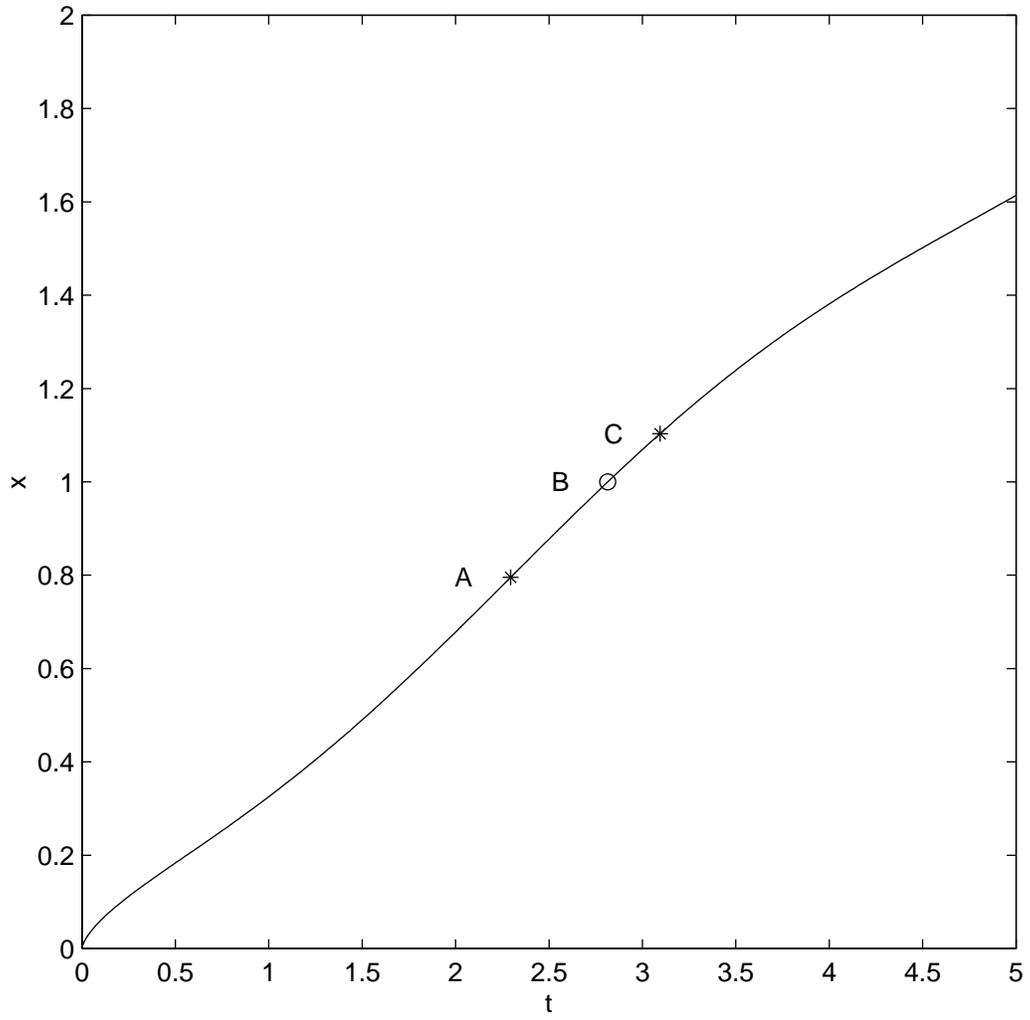
Figure 1.  Scale factor $x(t)$; point A marks the first switch from 
acceleration to deceleration; $q_0=0.5$ at point B, and the Hubble time
exceeds the age after point C.
\vskip 0.3cm

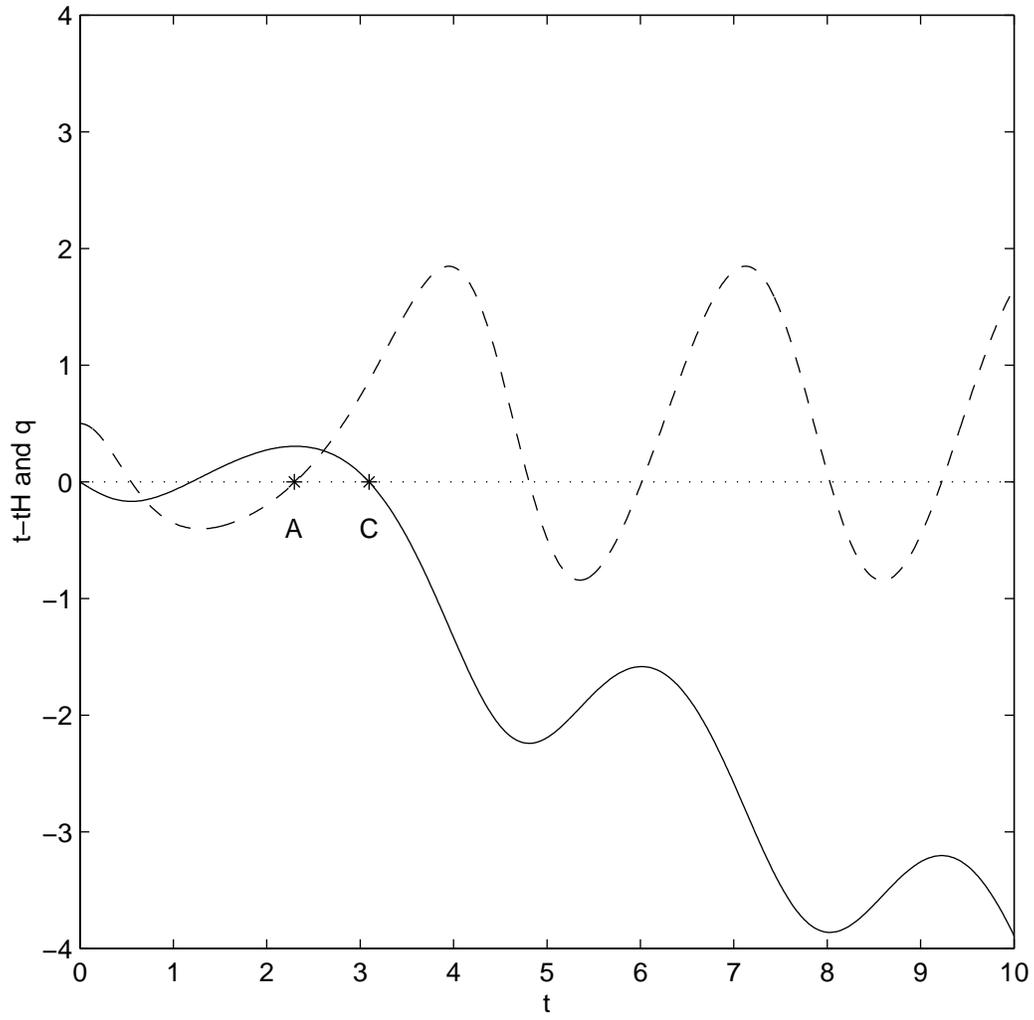
Figure 2.  Continuous line shows $t-t_H$ as a function of $t$, and the
dashed line shows the deceleration parameter $q(t)$; in the region A to C
we have $t>t_H$ and $q>0$.
\vskip 0.3cm

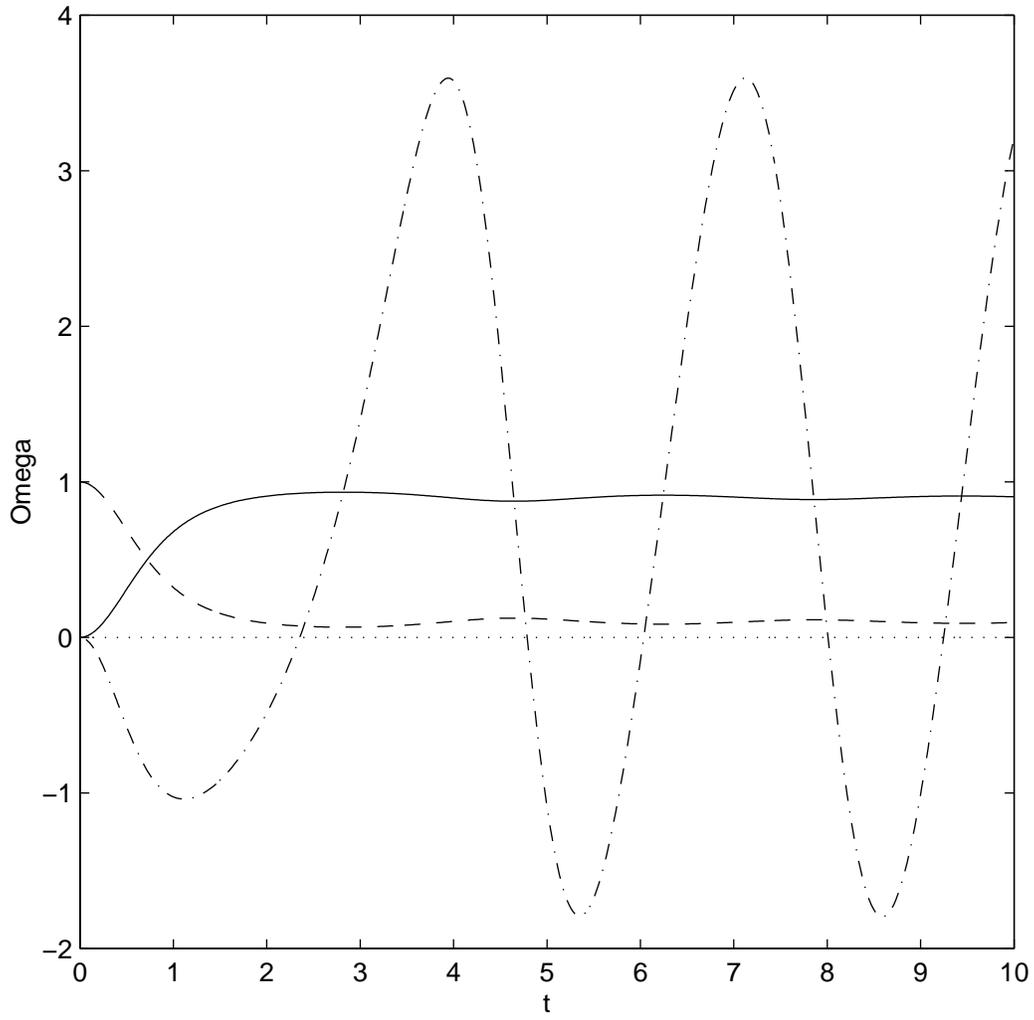
Figure 3.  $\Omega_{\rho_\phi}$ (continuous), $\Omega_m$ (dashed) 
and $\Omega_{\rho_\phi}+3\Omega_{p_\phi}$ (dash-dotted) as functions of $t$. 
\vskip 0.3cm

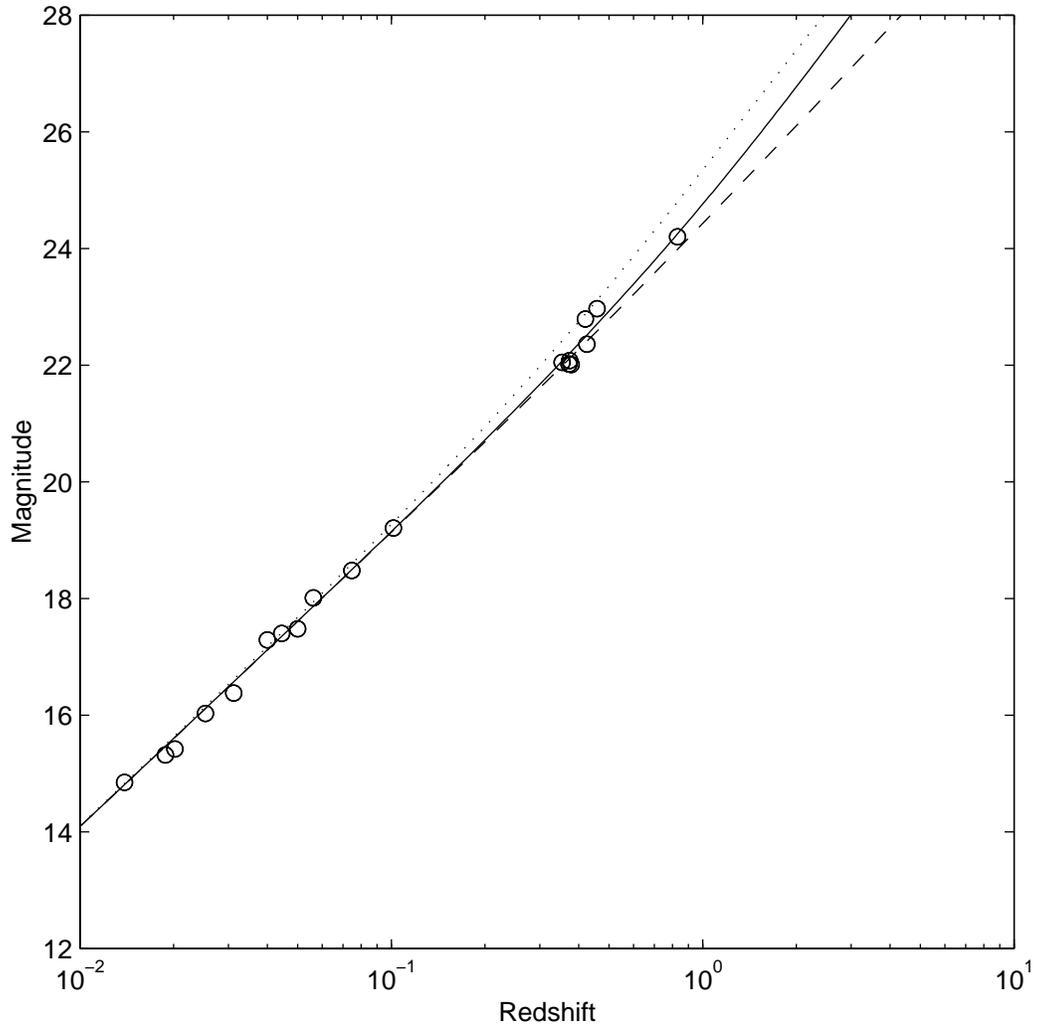
Figure 4.  Magnitude-redshift relation for the scalar field model discussed 
in Section 2 (continuous line), for the canonical $\Omega_0=1$, $q_0=0.5$ 
matter-dominated model (dashed), and for a flat accelerating model
$\Omega_0=0.1$, $\Lambda_0=0.9$ (dotted).  Data points for Type 1a supernovae 
due to Perlmutter at al. (1997b) and Hamuy et al. (1995).
\vskip 0.3cm

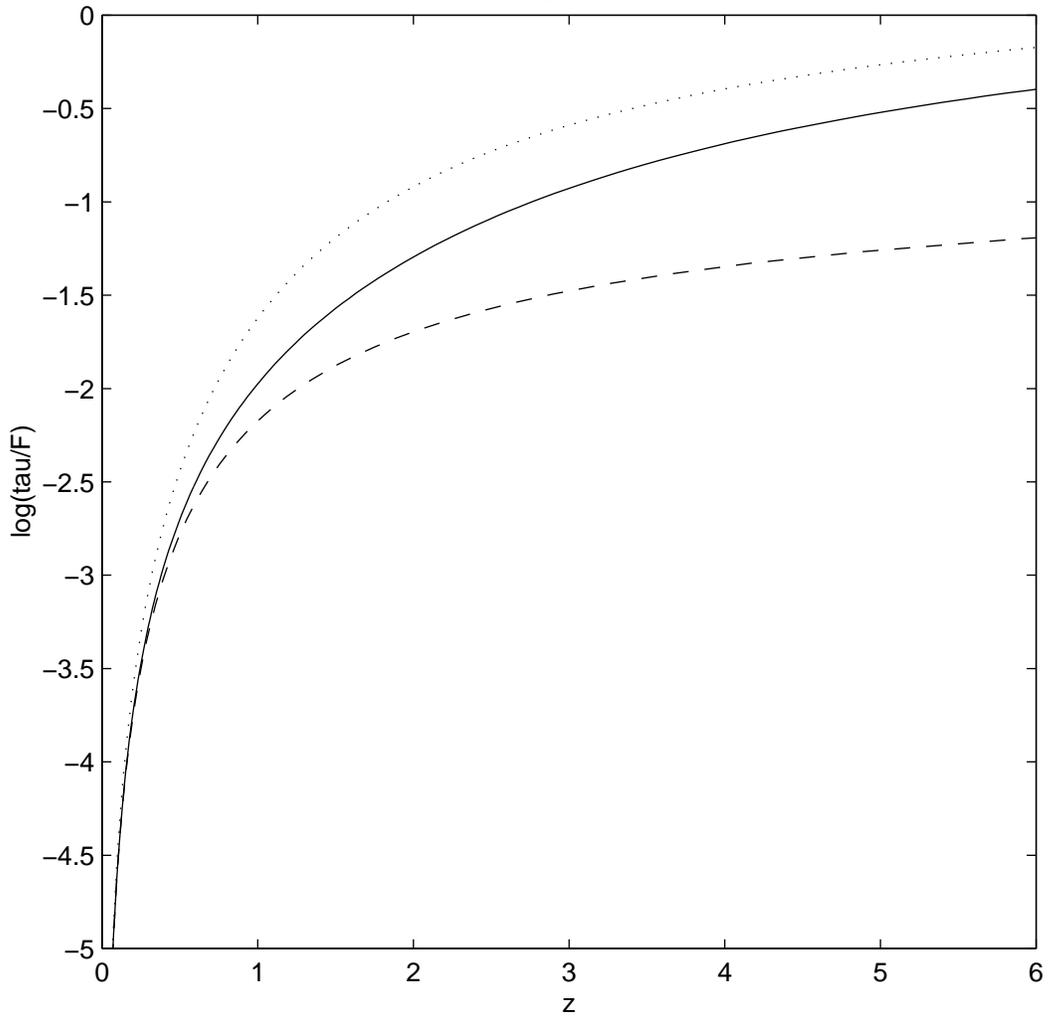
Figure 5.  Gravitational lens optical depth as a function of redshift
for the scalar field model discussed in Section 2 (continuous line), 
for the canonical $\Omega_0=1$, $q_0=0.5$ matter-dominated model (dashed), 
and for a flat accelerating model $\Omega_0=0.1$, $\Lambda_0=0.9$ (dotted).  

\bye